 \documentclass[useAMS,a4paper,fleqn,usenatbib]{mnras}
 \usepackage{times,graphicx,amssymb,amsmath,natbib}

\newcommand{\msun}{\,\mathrm{M}_{\sun}}
\newcommand{\mpc}{\,\mathrm{Mpc}\,h^{-1}}

\title[N/O in galaxies from chemodynamical simulations]{Evolution of N/O ratios in galaxies from cosmological hydrodynamical simulations}
\author[F. Vincenzo \& C. Kobayashi]{Fiorenzo Vincenzo$^{1}$\thanks{f.vincenzo@herts.ac.uk} \& 
Chiaki Kobayashi$^{1}$\thanks{c.kobayashi@herts.ac.uk} 
\\
$^{1}$Centre for Astrophysics Research, University of Hertfordshire, College Lane, Hatfield, AL10 9AB, UK }

\begin{document}

\date{Accepted 2018 April 20. Received 2018 March 20; in original form 2018 January 22}

\pagerange{\pageref{firstpage}--\pageref{lastpage}} \pubyear{2018}

\maketitle

\label{firstpage}


\begin{abstract}
We study the redshift evolution of the gas-phase O/H and N/O abundances, 
both (i) for individual ISM regions within single spatially-resolved galaxies and (ii) when dealing with average abundances 
in the whole ISM of many unresolved galaxies. 
We make use of a cosmological hydrodynamical 
simulation including detailed chemical enrichment, which properly takes into account the variety of different 
stellar nucleosynthetic sources of O and N in galaxies. 
We identify $33$ galaxies in the simulation, 
lying within dark matter halos with virial mass in the range $10^{11}\le M_{\text{DM}} \le 10^{13}\,\text{M}_{\sun}$ and 
reconstruct how they evolved with redshift. 
For the local and global measurements, the observed increasing trend of N/O at high O/H can be explained, respectively, (i) as the consequence of metallicity gradients 
which have settled in the galaxy interstellar medium, where the innermost galactic regions have the highest O/H abundances and 
the highest N/O ratios, and (ii) as the consequence of an underlying average mass-metallicity relation 
that galaxies obey as they evolve across cosmic epochs, where -- at any redshift -- less massive galaxies have 
lower average O/H and N/O ratios than the more massive ones. 
We do not find a strong dependence on the environment. 
For both local and global relations, 
the predicted N/O--O/H relation is due to the mostly secondary origin of N in stars. 
We also predict that the O/H and N/O gradients in the galaxy interstellar medium 
gradually flatten as functions of redshift, with the average N/O ratios being strictly coupled with 
the galaxy star formation history. Because N production strongly depends on O abundances, 
we obtain a universal relation for the N/O--O/H abundance diagram whether we consider average 
abundances of many unresolved galaxies put together or many abundance measurements within a single spatially-resolved galaxy.
\end{abstract}




\section{Introduction} \label{sec:intro}

{Elemental abundances are widely used in astrophysics to constrain the star formation history (SFH) of galaxies 
(e.g., \citealt{kobayashi2016}). An example of a SFH chemical abundance diagnostic is 
given by [$\alpha$/Fe]\footnote{By $\alpha$-elements 
we usually mean O, Mg, Ne, Si, S, Ca. The square bracket notation for the 
 stellar chemical abundances is defined as follows: $[X/Y] = \log(N_{X}/N_{Y})_{\star} - \log(N_{X}/N_{Y})_{\odot}$, where 
 $N_{X}$ and $N_{Y}$ represent the number density of the chemical elements $X$ and $Y$, respectively. }; 
 from the observed [$\alpha$/Fe]--[Fe/H] relations, chemical evolution models have demonstrated 
 that the various constituents of our Galaxy (halo, bulge, thick and thin disc) formed on different typical time scales 
 (see, for example, \citealt{chiappini1997,grieco2012,brusadin2013,micali2013,spitoni2016,grisoni2017}); 
 furthermore, by making use of the [$\alpha$/Fe] ratio estimated 
 from spectral indices, chemical evolution models have depicted early-type elliptical galaxies as forming from a short and 
 intense burst of star formation in the past, in agreement with observations \citep{matteucci1994,thomas2003,pipino2004,taylor2015a,taylor2015b,kriek2016,demasi2018}. 
 The observed [$\alpha$/Fe]--[Fe/H] diagram can be effectively used as a SFH diagnostic \textit{(i)} firstly, 
 because $\alpha$-elements and Fe are mostly released on different 
 typical time scales by core-collapse and Type Ia Supernovae (SNe), respectively, and 
 \textit{(ii)} secondly, because the nucleosynthesis of 
 $\alpha$-elements in stars is not correlated with their Fe abundance \citep{kobayashi2006}. }
 
In the star forming disc galaxies, however, it is not possible to measure iron abundances. 
For this reason, the O/H elemental abundance as well as the N/O\footnote{For brevity, we use the following notation: $\text{N/O}\equiv\log({\text{N/O}})_{\text{gas}}$ and 
$\text{O/H}\equiv\log({\text{O/H}})_{\text{gas}}+12$, for the gas-phase chemical abundances.} abundance ratio are among the most measured metallicity proxies 
in the interstellar medium (ISM). 
Current Galactic and extragalactic spectroscopic surveys such as MaNGa \citep{bundy2015} 
are capable of reaching resolutions which were unimaginable only 
few decades ago. Large amounts of observational data are constantly being released, challenging theorists to develop models 
which can explain at the same time the variety of different physical observables nowadays available. 
One of the most important pieces of information we can extract 
from extragalactic spectroscopic surveys is 
the N/O--O/H diagram.

Historically,  the N and O abundances have been measured for individual targets (e.g. \textsc{Hii} regions or star forming regions) within a number of nearby galaxies 
(e.g., \citealt{garnett1990,vilacostas1993,izotov1999,pilyugin2010,berg2016,magrini2017}); then large-scale spectroscopic surveys have improved the 
statistics considerably, where chemical abundances have been measured from the 
integrated galaxy spectra (e.g. \citealt{andrews2013}, and references therein); finally, thanks to multi-object spectrographs and 
integral field unit (IFU) surveys, now it is possible to resolve the abundance patterns in star forming regions 
within a large number of galaxies \citep{perezmontero2016,sanchez-menguiano2016,belfiore2017}. 
This growing amount of observational data 
has suggested the use of the N/O--O/H relation as an alternative 
SFH chemical abundance diagnostic of galaxies \citep{chiappini2005,molla2006,vincenzo2016a}.
In near future, it will be possible to obtain these elemental abundances in high-redshift galaxies with JWST/NIRSpec and study the redshift evolution of the N/O--O/H relation.

Before studying the redshift evolution, it is important to understand the origin of the observed N/O--O/H relations in the local Universe.
The observed relations have been obtained with O/H and N/O abundance measurements both 
 \textit{(i)} as \textit{global} average values, measured 
 from the galaxy integrated spectra and hence representative of unresolved galaxies, and 
 \textit{(ii)} as \textit{local} abundance measurements in resolved \textsc{Hii} or 
star forming regions within single, spatially-resolved external galaxies; 
these two cases are conceptually different with respect to each other and may give rise -- in principle -- to different N/O--O/H relations. 

All the chemical elements with atomic number $A\ge12$ in the cosmos are synthesised in stellar interiors either during the 
quiescent phases of hydrostatic burning or through explosive nucleosynthesis during SN explosions \citep{arnett1996}. 
If a theoretical model is to make predictions about the chemical abundances 
coming from the analysis of stellar spectra, the chemical enrichment feedback from star formation activity must be properly included in the 
theoretical machinery by taking into account the 
variety of different nucleosynthesis sources which can actually produce a given chemical element $X$; the different distributions 
of delay times between the formation of each astrophysical source and its death must also be taken into account 
(see \citealt{matteucci2001,matteucci2012,pagel2009} for exhaustive reviews on the subject). 

Detailed chemical evolution of galaxies have mostly been studied by making use of one-zone models (e.g. \citealt{henry2000,chiappini2005,vincenzo2016a}, but 
see also \citealt{vangioni2017}, where a one-zone model has been embedded in a cosmological framework), which are based on 
the so-called instantaneous mixing approximation. However, in a real galaxy, chemical enrichment is inhomogeneous, which is 
important if we want to constrain the SFH from $X/Y$ abundance ratio diagrams \citep{kobayashi2011a}. 
Cosmological chemodynamical simulations are nowadays the best tools to shed light 
on how the SFH took place in different galaxies. These simulations are
also key to understanding how chemical elements are synthesised, released and 
later distributed within galaxies, because they are  able 
to address the large amounts of data 
which are already available or about to come. In fact, cosmological chemodynamical simulations can provide a unifying 
picture for the formation and evolution of the many different populations of galaxies in the Universe (see, for example, 
\citealt{maio2015}). 

An advantage of using chemodynamical simulations is that one can predict both local and global relations for a large sample of 
simulated galaxies; another advantage is that  one can naturally have chemical abundance gradients as functions of the 
galactocentric distance within the ISM of the simulated galaxies. 
By using chemical abundance measurements from the Cepheids \citep{andrievsky2002,luck2003,lucklambert2011,korotin2014,genovali2015}, 
planetary nebulae \citep{maciel1994,costa2004,stanghellini2006,gutenkunst2008} or \textsc{Hii} regions \citep{deharveng2000,esteban2005,rudolph2006,fernandez2017,esteban2017} 
many observational works  have shown, for example, that O/H in our Galaxy 
steadily diminishes when moving outwards as a function of the galactocentric distance, but radial gradients have been observed by those works also 
for other chemical elements; furthermore, \citet{belfiore2017} have shown that 
the N/O ratios can vary as functions of both the galactocentric distance and stellar mass, when considering a 
large sample of nearby galaxies in the MaNGa survey. 

Historically, multi-zone chemical evolution models have been constructed to reproduce the observed radial metallicity gradients in the Galactic disc 
by assuming the so-called ``inside-out scenario'', according to which the innermost (most metal-rich) Galactic regions assembled on much shorter typical timescales 
than the outermost (most metal-poor) ones, namely by assuming that the Galaxy formed \textit{from the inside out} (see, for example, \citealt{chiappini2001,cescutti2007,magrini2009,spitoni2011}). 
Chemical evolution models with inside-out growth of the disc 
and the star formation efficiency being modulated by the angular velocity of the gas 
predict a flattening of the radial metallicity gradients as a function of time  \citep{portinari1999,boissier2000}; also chemodynamical simulations 
usually predict a flattening of the radial metallicity gradients as a function of time (see \citealt{kobayashi2011a,pilkington2012,gibson2013} 
and references therein). 
Finally, there are chemical evolution models predicting an inversion of the radial metallicity gradients at high redshift, corroborated by some observational findings 
(see \citealt{cresci2010,werk2010,queyrel2012,mott2013}, but also \citealt{schoenrich2017} for a critical discussion).

In this work, we show the results of our cosmological chemodynamical simulation 
including the latest stellar nucleosynthesis yields; 
we investigate both \textit{(i)} local and \textit{(ii)} global 
N/O--O/H relations, i.e. \textit{(i)} the relations obtained of individual targets 
\textit{within} single spatially-resolved galaxies and 
\textit{(ii)} the relations obtained with average abundances for the whole ISM of many unresolved galaxies put together. 
If the predicted relations follow a similar trend in the N/O--O/H diagram, we try to understand the causes of this in galaxies. 
 Moreover, we show our predictions for the redshift evolution of the O/H and N/O radial gradients of a sample of 
 galaxies in our cosmological simulation; finally, we show how the simulated galaxies 
 move in the N/O--O/H, stellar mass--O/H and stellar mass--N/O diagrams 
 as they evolve across cosmic epochs, fully exploiting the predictive power of a cosmological hydrodynamical simulation. 
We would like to note again that only by making use of
  chemodynamical simulations can we study both {\it local} and {\it global} evolution of chemical abundances, and that cosmological simulations allow us to study the effect of environment on the chemical evolution of galaxies as well.
 
 Our work is organised as follows. In Section \ref{sec:model} we summarise the main assumptions of our model and the analysis method of the 
 simulation. In Section \ref{sec:results} we present the results of our study. 
 We first discuss the origin of the local and global N/O--O/H
  relations for nearby galaxies, and then show the redshift evolution
  and the environmental dependence in Section \ref{sec:cosmological_context}. 
Finally, in Section \ref{sec:conclusions} we draw our conclusions.

\section{Simulation model and methods} \label{sec:model}
Our simulation code is based on the \textsc{Gadget-3} code \citep{springel2005} and relevant baryon physics is included, namely 
UV background heating, metal-dependent radiative cooling, star formation, thermal stellar feedback, and chemical enrichment from asymptotic giant branch 
(AGB) stars, 
core-collapse and Type Ia supernovae (SNe). 
 Therefore, the star formation activity  
 within the ISM of galaxies is affected both by the thermal energetic feedback and by the chemical enrichment 
 of star particles through stellar winds and SN explosions 
 (see \citealt{kobayashi2004,kobayashi2007,kobayashi2011a,taylor2014} 
 for a detailed description of the model).

 In summary, 
 we evolve a cubic volume of the standard $\Lambda$-cold dark matter Universe with side $10\mpc$, 
 periodic boundary conditions, and the cosmological 
 parameters being given by the nine-year Wilkinson Microwave Anisotropy Probe \citep{hinshaw2013};
$\Omega_0=0.28$, $\Omega_{\Lambda}=0.72$, $\Omega_{\mathrm{b}}=0.046$,
$H_0=100\times h=70$\,km\,$^{-1}$\,Mpc, and $\sigma_8 = 0.82$. 
 The mass resolution of our simulation is $M_{\mathrm{DM}} \approx3.097\times10^{7}\,h^{-1}\msun$ for the dark matter (DM) component 
 and $M_{\mathrm{gas}}=6.09\times10^{6}\,h^{-1}\msun$ for the gas fluid.  Finally, in our simulation we assume a gravitational 
softening length $\epsilon_{\mathrm{gas}}\approx0.84\,h^{-1}\;\mathrm{kpc}$, in comoving units. 

{ The initial conditions of our simulation are the same as in \citet{kobayashi2007}, but with updated cosmological parameters and 
better resolution; in particular, we assume initial conditions giving rise to a standard field at redshift $z=0$, with no strong central concentration of galaxies. Our initial conditions are different from those in \citet{taylor2014}.}

\begin{figure}
\centering
\includegraphics[width=8.0cm]{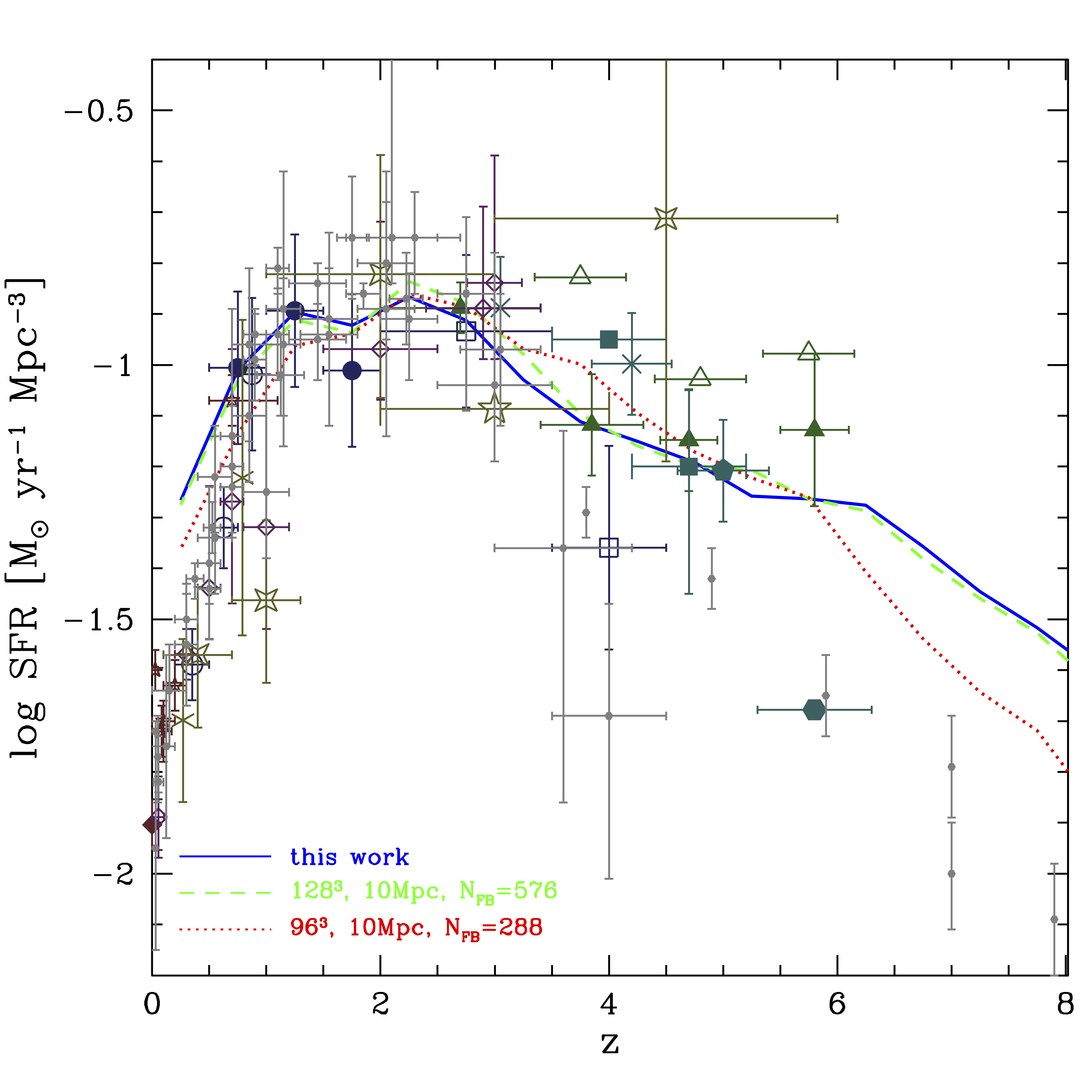} 
\caption{The predicted cosmic SFR as a function of redshift from different cosmological hydrodynamical simulations. The blue solid line corresponds 
to the predictions of the simulation of this work, with chemical enrichment including failed SNe; the green long dashed line to the same simulation as in this 
work but without failed SNe (original \citealt{kobayashi2011b} stellar yields); finally, the red dotted line corresponds to the prediction of a similar simulation. 
The observational data are taken from \citet[small grey points]{madau2014} and from \citet[see references therein]{kobayashi2007}.   }
\label{fig:csfr}
\end{figure}

 \subsection{Chemical enrichment model} \label{sec:chem_model}
 According to their mass and metallicity, stars at their death pollute the ISM of galaxies 
 with different fractions of a given chemical element. We cannot resolve single stars in our simulation, hence we assume 
 that each star particle represents a simple stellar population (SSP) with fixed age and chemical composition. 
 Then we assume that all the embedded stars within 
 each single SSP have a universal mass-spectrum at their birth which follows the \citet{kroupa2008} initial mass function (IMF), 
 as defined in the 
stellar mass range $0.01\le m \le 120\,\text{M}_{\sun}$. As each given SSP gets older and older as a function of 
cosmic time, embedded stars with lower and lower mass enrich the surrounding gas particles with their nucleosynthetic products; 
the number of dying stars within a given SSP at the time $t$ is given by the assumed IMF and SSP 
mass, while the enrichment time of a star with mass $m$ and metallicity $Z$ is given by the assumed stellar lifetimes, 
$\tau(m,Z)$; in this work, we assume the stellar lifetimes of 
\citet{kobayashi2004}, which are both metallicity- and mass-dependent.

In our simulation, the stellar nucleosynthetic yields are the same as in \citet{kobayashi2011b}, which include the chemical enrichment of 
AGB stars and SN explosions. 
The effect of hypernovae is included in our simulation for stars with mass $m \ge 25\,\text{M}_{\sun}$ 
{ with the following metallicity-dependent hypernova fraction: $\epsilon_{\rm HN}=0.5, 0.5, 0.4, 0.01$, and $0.01$ for 
$Z=0, 0.001, 0.004, 0.02$, and $0.05$, respectively, which is necessary to match the observed elemental abundances in the Milky Way \citep{kobayashi2011a}.} 
We additionally assume that all stars with mass $m \ge 25\,\text{M}_{\sun}$ and metallicity $Z \ge 0.02$ {which are not hypernovae} 
end up their lives as failed SNe \citep{smartt2009,muller2016} 
and pollute the galaxy ISM only with H, He, C, N and F, which are synthesised in the outermost shells of the 
SN ejecta; the other chemical elements (including O) are assumed to fall back into the black hole, hence they are not 
expelled by the star into the surrounding ISM 
(see also \citealt{vincenzo2018}; Kobayashi et al., in prep.). 

We assume that each galaxy SSP distributes thermal energy and stellar nucleosynthetic products 
to its closest $576$ neighbour gas particles (with the smoothing kernel weighting). 
This value, together with the other parameters specified above, is chosen to match the observed cosmic star formation rate 
(SFR; \citealt{hopkins2006,madau2014}). 
Figure \ref{fig:csfr} shows the predicted cosmic SFR history of our simulation with failed SNe (blue solid line), as compared to the same 
simulation but with the original yields from \citet[green long dashed line]{kobayashi2011b} without failed SNe. 
There is no significant difference in the cosmic SFRs and in the basic properties of the galaxies such as mass and morphology. 
The red dotted line in Figure \ref{fig:csfr} shows the predictions for the cosmic SFR of a 
similar simulation but with lower resolution ($2 \times 96^{3}$ particles) than in this work ($2 \times 128^{3}$ particles); 
an agreement between the two can be only found by assuming a different number of feedback neighbour particles, $N_{\text{FB}}$. 
In particular, by increasing the resolution and keeping $N_{\text{FB}}$  constant, ISM regions with higher and higher densities 
can be resolved and the SN feedback 
affects smaller regions around each given star particle; therefore, to obtain similar results in simulations 
with higher resolution, $N_{\text{FB}}$ should be increased accordingly \citep{kobayashi2007}. 

   \begin{figure}
\centering
\includegraphics[width=6.0cm]{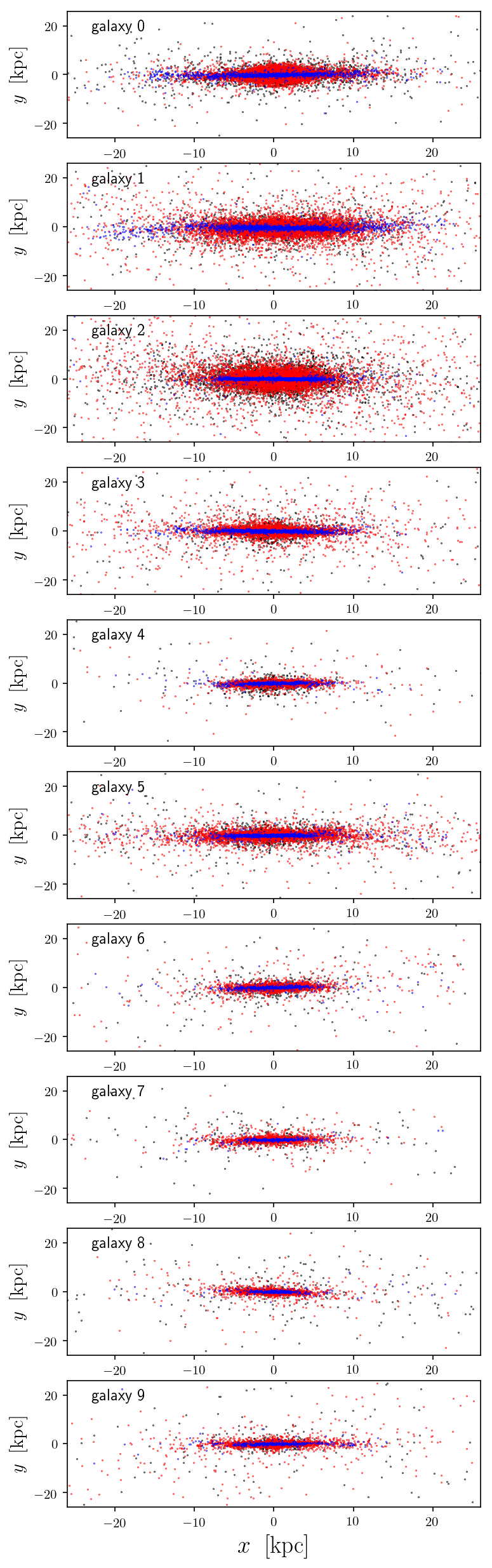} 
\caption{Our ten reference galaxies when viewed edge-on. Black points correspond to the older stellar populations ($>90$ per cent in 
the cumulative age distribution function); blue points to younger star particles ($<10$ per cent in 
the cumulative age distribution function); finally, red points correspond to intermediate-age stellar populations 
which lie between $10$ and $90$ per cent 
in the cumulative age distribution function.   }
\label{fig:gal_im_star}
\end{figure}

{ Although our resolution is good enough to study radial gradients of chemical abundances in galaxies, it is not possible to resolve 
the small-scale physics within star-forming clouds and SN ejecta in galaxy simulations; 
for this reason, chemical enrichment is included by computing the 
contribution from each single star particle, depending on the metallicity. Therefore the chemical feedback 
can vary as a function of time and location within the galaxy \citep{kobayashi2004,kobayashi2011a}. 
We remark on the fact that the evolution of the elemental abundance ratios in the ISM 
is mainly driven by the difference in the age and metallicity of the enrichment sources, being less affected by the uncertainty in the 
ISM metallicity due to the limited resolution. }

Since in this paper we focus on the evolution of the N/O ratio in galaxies, we briefly recall here how O and N are synthesised by stars in galaxies (see also 
\citealt{kobayashi2011b,vincenzo2016a} for more details). 
First of all, both N and O can be produced by massive stars, with mass $m > 8\,\text{M}_{\odot}$, dying as core-collapse SNe on 
short typical time scales after the star formation event 
($\lesssim30\,\text{Myr}$); in this case, stellar evolutionary models predict N to be mainly produced as a secondary element in massive stars,  
in the CNO cycle at the expense of C and O nuclei already present in the gas mixture at the stellar birth. 
One-zone chemical evolution models showed  that 
on its own, the ``secondary'' N component from massive stars is not sufficient to reproduce the observed N/O plateau in our Galaxy 
at very low metallicity \citep{matteucci1986,chiappini2005,chiappini2008}. Therefore, following 
the original suggestion of  \citet{matteucci1986}, many one-zone models assumed 
an additional primary N production by massive stars to reproduce the observed N/O plateau 
at very low metallicity \citep{pettini2002,pettini2008,spite2005,pilyugin2010}, which is however highly scattered. 
In our simulation, we do not assume 
any additional primary N production for massive stars.

Low- and intermediate-mass (LIM) 
stars, with mass in the range $4\lesssim m \lesssim 8 \text{M}_{\sun}$, 
are dominant stellar nucleosynthesis sources of N, when experiencing the 
AGB phase (see, for example, \citealt{ventura2013,ventura2017} 
for more details). Most of the nitrogen from AGB stars is secondary and its stellar yields steadily increase 
as functions of the initial stellar metallicity. Note that, however, there may be also a primary N component that can be important in the 
chemical evolution of galaxies at very low metallicity, which is predicted when hot-bottom burning occurs in 
conjunction with the so-called third dredge-up (see also \citealt{vincenzo2016a} and reference therein).

 \subsection{Analysis of the simulation}

 From our cosmological simulation, we create a catalogue of $33$ stellar systems at redshift $z=0$, 
 all embedded within dark matter (DM) halos with virial mass in the range 
 $10^{11} \le M_{\text{DM}} \le 10^{13}\,\text{M}_{\sun}$; we make use of the \textsc{Rockstar} 
 friend-of-friends group-finding algorithm with adaptive hierarchical refinement to determine all the DM halos in the 
 simulation \citep{behroozi2013}. The $33$ stellar systems of our catalogue span a variety of different star formation histories (SFHs) and 
 consequently show different chemical evolution histories from their formation epoch. For each stellar system at redshift $z=0$ in our catalogue, 
 we retrieve the main physical and chemical properties of all its star and gas particles 
 going back in redshift, by means of a simple searching algorithm (each particle in the simulation is univocally 
 characterised by an ID number). At all redshifts, each galaxy in our catalogue is defined as follows.

\begin{enumerate} 
 
\item  At any given time $t_{1}$ in the past, we fit with Gaussian functions the normalised density-weighted distributions of the $x$, $y$ and $z$ 
 coordinates of all the gas particles within the galaxy, which have
 been retrieved from the simulation snapshot at a time shortly after
 $t_{2}=t_{1}+\Delta t$; then we consider in our 
 analysis all the star and gas particles at the time $t_1$ in the simulation 
 which lie within $4\sigma$ from the centre of the three Gaussians. Therefore, in the presence 
 of merger events, we choose to follow the stellar system with the highest gas densities. 
 
\item If the fitting procedure fails at a given redshift (usually 
    corresponding to high velocity encounters or minor/major mergers), we broaden 
 our criteria and consider at that redshift all the 
 gas and star particles which lie within $20\;\text{kpc}$ from the centre of mass of the star particles 
 which have been retrieved from the subsequent simulation time step. 
 
  \end{enumerate}
 \noindent By following the analysis as described above, we can study the evolution of the galaxy physical properties continuously as functions 
 of redshift with an automated algorithm. 
 The small fluctuations in the predicted evolution of the average galaxy properties are mostly due to an imperfect centring on the galaxy 
 main body, particularly associated with merging episodes with other stellar systems.

\section{Results} \label{sec:results}

   \begin{figure}
\centering
\includegraphics[width=6.0cm]{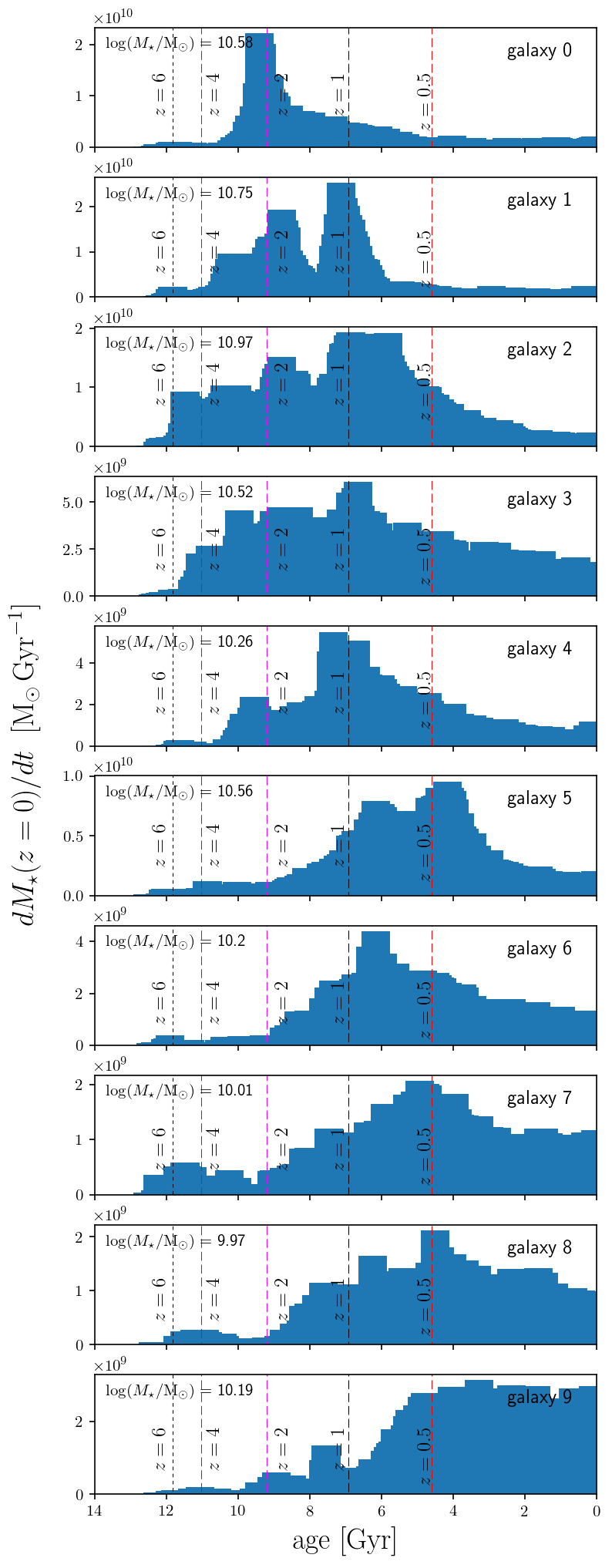} 
\caption{The stellar mass distribution function of our ten reference galaxies as functions of the stellar birth time. The area under the predicted distribution 
corresponds to the total galaxy stellar mass at redshift $z=0$. From top to bottom, our ten reference galaxies have decreasing average stellar ages. The figure can 
be read as the fraction of the present-day galaxy stellar mass coming from each given age bin in the past. In the upper corner on the left 
we report for each panel the total galaxy stellar mass at $z=0$. }
\label{fig:age_distribution}
\end{figure}

In this Section, we present our new results from the analysis of the cosmological hydrodynamical simulation
described in Section \ref{sec:model}.  
In this work, we select ten representative star forming disc galaxies
from our catalogue, so that 
these reference galaxies have a range of characteristic
SFHs. All of our ten reference galaxies lie within DM halos with 
virial masses in the range $10^{11} \le M_{\text{DM}} \le 10^{12}\,\text{M}_{\sun}$. In the first part of this Section, 
we show our predictions for the gas-phase O/H--N/O abundance patterns in our reference galaxies. In the second part, 
we show how the average O/H and N/O abundance ratios evolve with time when considering our entire 
sample of $33$ galaxies. 

\subsection{Star formation history of the reference disc galaxies}

Our ten reference galaxies are shown in Figure \ref{fig:gal_im_star}, as viewed 
edge-on. Different colours in Figure \ref{fig:gal_im_star} correspond to galaxy stellar populations with different age; 
in particular, the black points correspond to the older galaxy stellar populations 
($>90$ per cent in the cumulative age distribution function); blue points to younger star particles ($<10$ per cent in 
the cumulative age distribution function); finally, red points correspond to intermediate-age stellar populations which lie between $10$ and $90$ per cent 
in the cumulative age distribution function. 

In Figure \ref{fig:age_distribution}, we show the distribution of the total stellar mass of our ten reference galaxies at redshift $z=0$ 
as a function of the stellar age. When passing from Galaxy $0$ to Galaxy $9$, the stellar mass growth 
history becomes more concentrated towards later and later epochs. 
While most of the galaxies have smooth SFHs, Galaxy $1$ {undergoes} a major merger of two stellar systems 
with average ages peaking at $\approx9$ and $7\,\text{Gyr}$ ago. 
After the onset of star formation at $z\sim6$, Galaxies 2 and 3 had a relatively rapid increase of SFR, while the other galaxies had a slow start 
with very low SFR at $z<4$. Galaxy 9 is the youngest and 
  maintains a high SFR at the present epoch.

\subsection{O/H--N/O relations within single resolved galaxies}

\begin{figure}
\centering
\includegraphics[width=6.0cm]{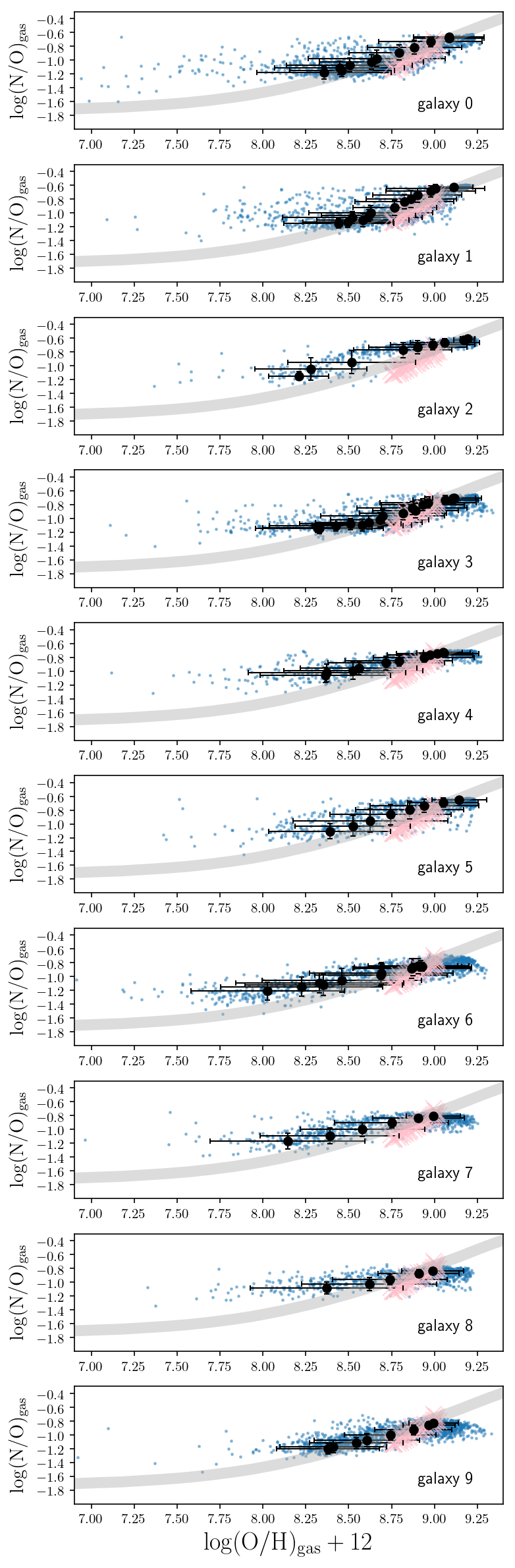} 
\caption{The predicted O/H--N/O abundance diagrams in our ten reference galaxies (blue points for each ISM location in the galaxy). 
The black points with the error bars correspond to the average O/H and N/O with the corresponding $1\sigma$ deviation, as derived by dividing our 
simulated galaxies in many concentric annuli and measuring the average gas-phase 
abundances within each annulus (in this case we only consider gas particles which lie within three times the galaxy half-mass radius, as computed 
from the stellar mass radial profile); 
blue points at higher O/H abundances lie in the inner galactic regions.  
We compare our simulation with the observed average relation of 
\citet[grey solid line; see references therein]{dopita2016} and the observed data 
from the MaNGa survey of \citet[pink symbols]{belfiore2017} for nearby disc galaxies with stellar 
mass in the range $9.5\le \log(M_{\star}/\text{M}_{\sun}) \le 11.0\,\text{dex}$.  }
\label{fig:no_oh}
\end{figure}

\begin{figure}
\centering
\includegraphics[width=6.0cm]{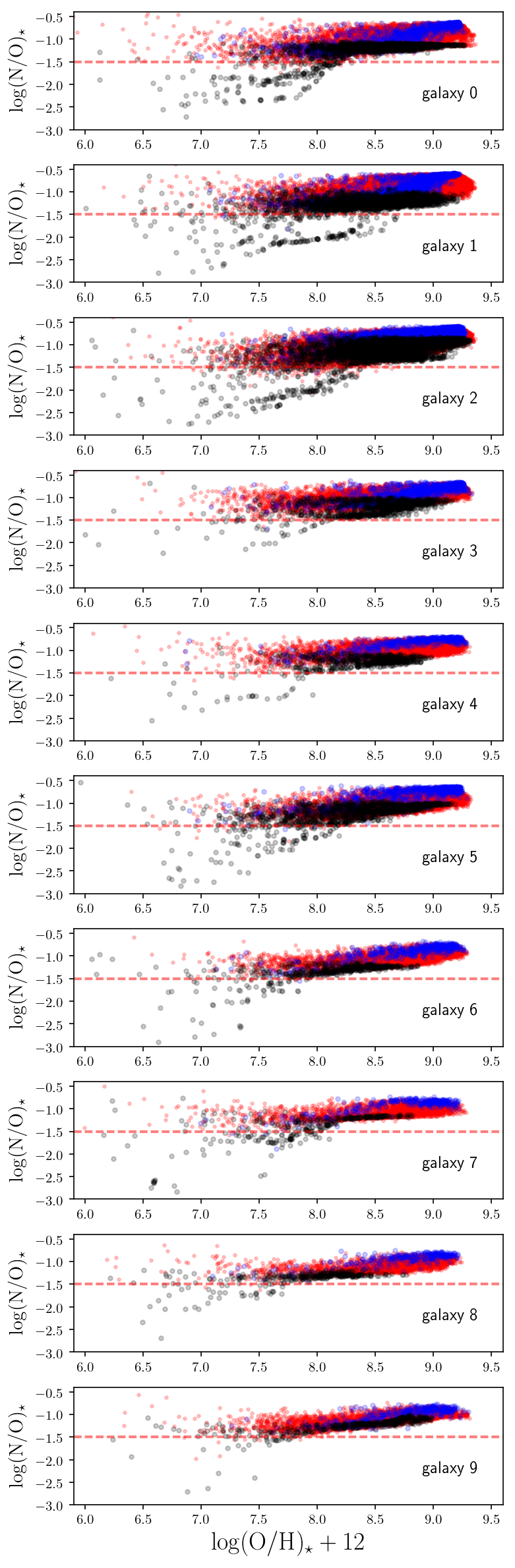} 
\caption{ {The predicted O/H--N/O abundance diagrams in the different stellar populations of our ten reference galaxies, from top to bottom. 
The black points represent the older stellar populations ($>90$ per cent in 
the cumulative age distribution function) in our reference galaxies; blue points to younger star particles ($<10$ per cent in 
the cumulative age distribution function); finally, red points correspond to intermediate-age stellar populations 
which lie between $10$ and $90$ per cent 
in the cumulative age distribution function. 
The red dashed line indicates the suggested average low-metallicity N/O-plateau at $\log(\mathrm{N/O})\approx-1.5\,\text{dex}$, as predicted by 
chemical evolution models with pure primary N production by massive stars. } }
\label{fig:no_oh_star}
\end{figure}

In Figure \ref{fig:no_oh}, the predicted gas-phase $\log(\text{O/H})$--$\log(\text{N/O})$ 
relation of individual ISM regions (blue points) in our ten reference
galaxies is compared 
to the observations in the local Universe \citep{dopita2016,belfiore2017}. 
We remark on the fact that the majority of the gas particles in our ten reference galaxies 
lie on a thin disc at the present time. 
The black points with the error bars represent 
the average O/H and N/O abundances with the corresponding $1\sigma$ 
standard deviation as predicted when dividing the galaxy in different concentric annuli of galactocentric distance 
and computing the average gas-phase O/H and N/O ratios therein; 
in this case, we only consider gas particles which lie within three times the galaxy half-mass radius, as computed from the stellar mass radial profile.  

The predicted N/O--O/H relation within all our reference galaxies qualitatively agrees 
with the observed N/O and O/H abundance measurements of \citet[pink symbols]{belfiore2017} 
from a large sample of spatially-resolved galaxies in the MaNGa survey. Our simulation also nicely follows the observed relation 
as derived by \citet[grey line]{dopita2016}, obtained by compiling
data from Milky Way metal-poor stars \citep{israelian2004,spite2005}, 
resolved \textsc{Hii} regions in blue compact galaxies \citep{izotov1999} and local B stars \citep{nieva2012}. 
At low metallicity, although we do not have many points, 
our predictions give slightly larger N/O, which may be due to the difference in the targets.
We note that the observed slopes of N/O vs. O/H of \citet{dopita2016}
and \citet{belfiore2017} differ from
each other because of well-known uncertainties in the abundance measurements, mostly due to the assumed calibration 
(see, for example, \citealt{kewley2008}, but also the discussion in \citealt{belfiore2017}).

In our simulations, the observed increasing trend of N/O vs. O/H 
is reproduced with failed SNe. 
If we do not assume failed SNe (namely, had we assumed the 
original \citealt{kobayashi2011b} stellar yield set), we predict almost flat trends and hence dramatically 
fail in reproducing the observed increasing trend of 
N/O vs. O/H at high metallicity (see \citealt{vincenzo2018}). 
On average, within all our reference galaxies in Figure \ref{fig:no_oh}, we predict that 
the galactic annuli with the largest O/H abundances 
lie in the innermost galactic regions and have also the largest N/O ratios; 
conversely, the galactic regions with the lowest O/H abundances are preferentially concentrated outwards and have the lowest N/O ratios. 

At redshift $z=0$, we have only a few low-metallicity components in the gas-phase of the galaxy ISM.
However, these are predicted to be common at higher redshift, at the
earliest evolutionary stages of the galaxies 
(see \citealt{vincenzo2018}). 
At redshift $z=0$, the low-metallicity components can be seen more clearly in the oldest galaxy stellar populations; 
this is shown in Figure \ref{fig:no_oh_star}, where the different galaxy stellar populations are discriminated with different 
colours in the N/O--O/H diagram according to their formation time, by
using the same criteria as in Figure \ref{fig:gal_im_star}.

At low metallicity, the majority of the older stellar population (black dots) show a flat trend as a function of metallicity.
We recall here that a flat trend of N/O in our chemodynamical model is caused by inhomogeneous chemical enrichment, 
where a significant contribution of AGB stars appears at low O/H. Depending on the relative
contribution between SNe and AGB stars, 
the exact value of the N/O ratios in the plateau may 
vary from galaxy to galaxy according to the galaxy formation time and SFH.  At $z=0$, this effect of inhomogeneous chemical enrichment 
is more important at the outskirts 
of our simulated galaxies because of the low SFRs (see \citealt{vincenzo2018}). 
On the one hand, massive galaxies 
that have relatively fast star formation also show very low N/O ratios, below the plateau, 
for the oldest and metal-poor stellar populations; these values 
are roughly in agreement with the observations in damped Ly$\alpha$ (DLA) systems \citep{pettini2002,pettini2008}, 
which are, however, also highly scattered \citep{zafar2014,vangioni2017}. 
On the other hand, the oldest and most metal-poor 
stellar populations in the less massive galaxies have $\log(\text{N/O})\sim-1.5\,\text{dex}$, on average.

 \begin{figure}
\centering
\includegraphics[width=8.5cm]{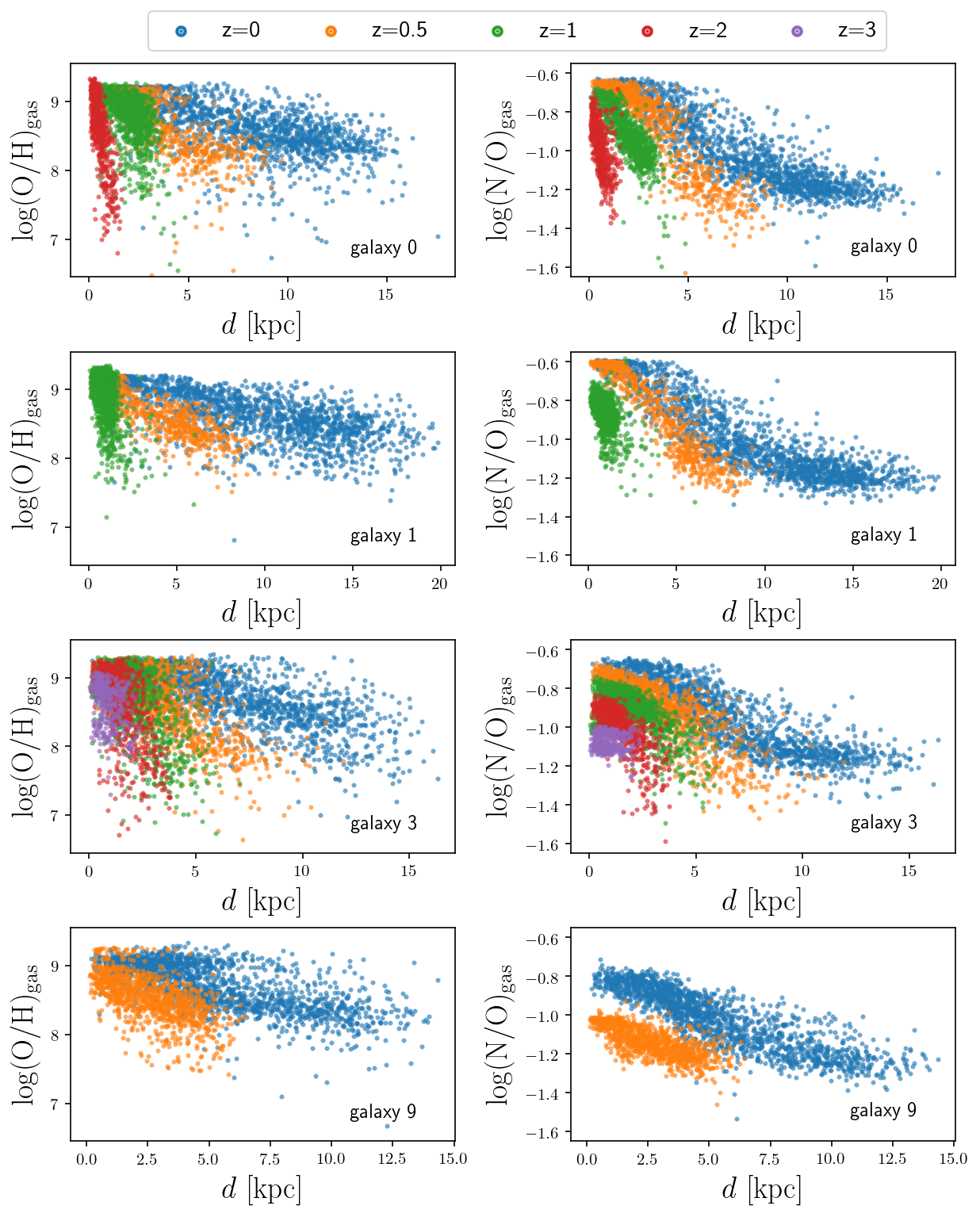} 
\caption{The predicted redshift evolution of the O/H (left panels) and N/O (right panels) gradients in the ISM of a sub-sample of four galaxies, 
as chosen from our ten reference galaxies. }
\label{fig:gradients}
\end{figure}

 \begin{figure*}
\centering
\includegraphics[width=14cm]{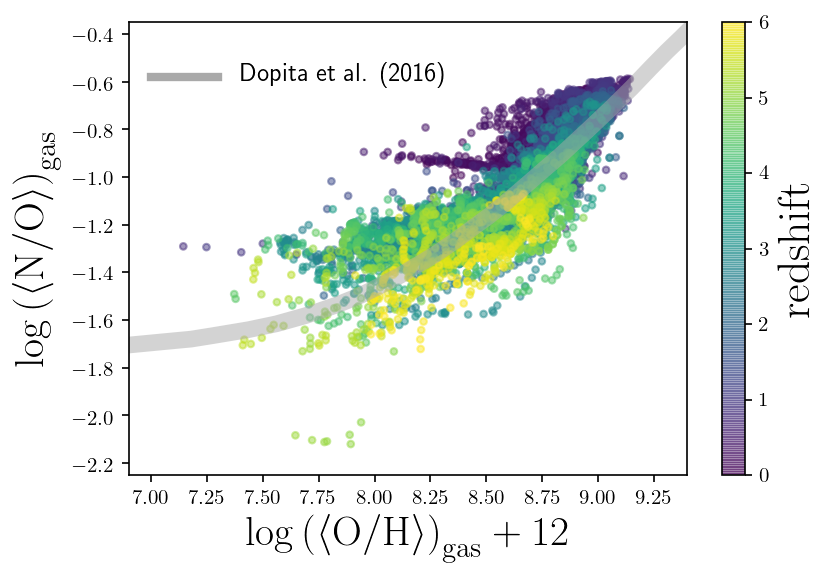}
\caption{The redshift evolutionary tracks of all $33$ galaxies in our catalogue in the O/H--N/O diagram. The abundances 
in the figure correspond to SFR-weighted averages in the gas-phase of
the ISM. The colour coding represents the redshift. }
\label{fig:no_oh_all}
\end{figure*}

In Figure \ref{fig:gradients}, we show how the gas-phase 
$\log(\text{O/H})$ and $\log(\text{N/O})$ ratios vary as functions of the galactocentric distance {and time} within a sub-sample 
of our ten reference galaxies. Different colours correspond to different redshifts when the gradients are computed. 
Although there is a dispersion in the chemical abundances at any fixed galactocentric distance (particularly 
for the O/H abundances; see also the horizontal error bars in Figure \ref{fig:no_oh}), 
after the first star formation episode in the galaxy, 
we predict a flattening of the abundance gradients 
as a function of time, together with an inside-out growth of the galaxy disc. 
In particular, in the very early ``protogalactic'' evolutionary stages, we predict highly scattered and overall flat abundance gradients; 
then, as the first series of 
stellar populations form and the galaxy contextually accretes gas from the environment, steep gas-phase abundance gradients 
develop, which then gradually flatten as a function of time. 
The predicted flattening of the abundance gradients with time is in agreement with the predictions of previous chemodynamical 
simulations (e.g., \citealt{pilkington2012}). 
Finally, the average O/H abundances at the centre do not show a strong redshift evolution, which is consistent with the observations in our Galactic bulge  \citep{zoccali2008}.

To explain our prediction of negative radial N/O gradients at redshift $z=0$, we recall here that the main N-producers in galaxies 
are not 
low-mass stars (see also Section \ref{sec:chem_model}), with chemical composition reflecting the chemical abundances in the ISM, 
from which they were born quite recently. 
The fact that we predict negative radial gas-phase O/H gradients at $z=0$ makes 
the N-producers more metal-rich inside than outside. 
Since N is mainly synthesised as a secondary element (namely, the N stellar yields 
increase, on average, as functions of the stellar metallicity), there is a correspondent 
increase of the average gas-phase N/O ratios 
when moving towards the inner galactic regions along the disc, where we predict the largest metallicities. 
In summary, the {\it local} O/H--N/O relation in Figure
\ref{fig:no_oh} can be explained as the consequence of radial
gradients in the disc within the galaxy ISM, as 
shown in Figure \ref{fig:gradients}.

\subsection{Global average N/O--O/H relation} \label{sec:globalNO_OH}

In this Section, {we show how all $33$ galaxies of our catalogue move in the N/O--O/H, 
mass--O/H and mass--N/O diagrams as a function of their evolutionary time. 
We focus on the average SFR-weighted $\log(\text{O/H})$ and $\log(\text{N/O})$ 
ratios of the whole ISM in the galaxies. } 

The main results of our analysis are shown in Figure \ref{fig:no_oh_all}, where we have put 
together all $33$ galaxies in our catalogue to show how they evolve in the N/O--O/H diagram; 
each point represents the SFR-weighted average $\log(\text{O/H})$ and $\log(\text{N/O})$ ratios in the ISM of each galaxy, 
and the colour coding indicates the redshift of the galaxy.  We find that 
the galaxies in our catalogue follow tracks in the N/O--O/H diagram which agree with the average 
\citet[solid grey line]{dopita2016} relation. 
Although we do not have many points at very low average gas-phase O/H abundances in our reference galaxies at high redshift, 
the points around $\log(\text{O/H}) \sim 7.5\,\text{dex}$ are in
  good agreement with the observations in DLA systems 
\citep{pettini2002,pettini2008,zafar2014}, in the halo stars of our Galaxy \citep{matteucci1986,spite2005}, 
and in irregular dwarf galaxies (e.g. \citealt{berg2016}), 
which exhibit $\log(\text{N/O})\approx-1.5\,\text{dex}$ with a large scatter around this value (see also \citealt{vincenzo2018}). 

\begin{figure}
\centering
\includegraphics[width=8.0cm]{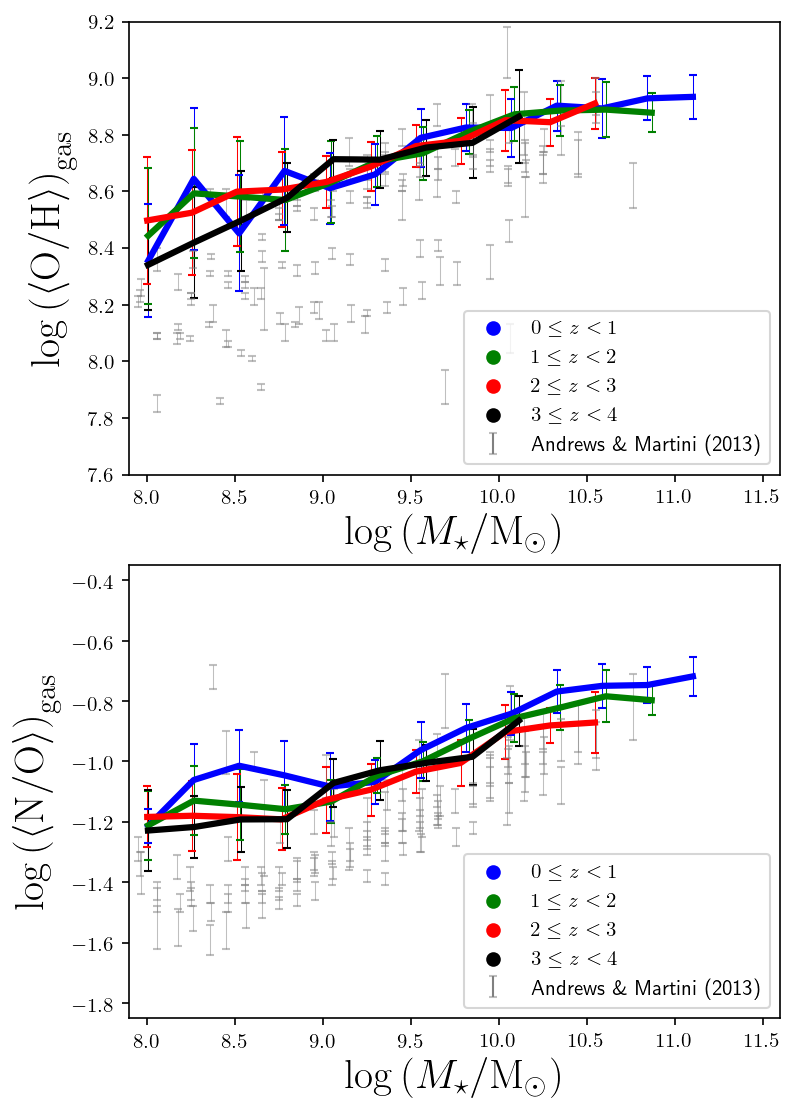} 
\caption{The redshift evolutionary tracks of all $33$ galaxies in our catalogue in the stellar mass--O/H (top panel) and 
stellar mass--N/O (bottom panel) diagrams. As in Figure \ref{fig:no_oh_all}, the abundances 
in the figure correspond to SFR-weighted averages in the gas-phase of the ISM. Different colours correspond to predicted 
mass--O/H and mass--N/O relations at different redshift intervals, as specified in the legends of the figure. The solid lines correspond to the 
average relation when dividing our catalogue in different stellar mass bins, 
while the error bars correspond to the $1\sigma$ standard deviation. { The grey points with error bars in both panels correspond to the 
average observed abundance measurements of \citet{andrews2013} in the redshift range $0.027 < z < 0.25$ from the SDSS DR7 \citep{abazajian2009}.  }
}
\label{fig:mass_met}
\end{figure}

In a given redshift interval, we predict that the highest N/O and O/H ratios are seen in the most massive galaxies.
Therefore, the {\it global} N/O--O/H relation can be explained as being determined by the galaxy stellar mass. To demonstrated this, 
in Figure \ref{fig:mass_met} we show how all $33$ galaxies of our catalogue evolve 
in the $\log(M_{\star}/\text{M}_{\sun})$--$\log\big( \langle \text{O/H} \rangle \big)$ diagram (top panel) and 
$\log(M_{\star}/\text{M}_{\sun})$--$\log\big( \langle \text{N/O} \rangle \big)$ diagram (bottom panel). Although there is 
a dispersion in the predicted average abundances, we predict well-defined correlations in the 
stellar mass--O/H and stellar mass--N/O diagrams in all redshift intervals. 
Compared with observations, 
our O/H and N/O abundances are systematically 
larger than the observed abundance measurements of \citet[grey points with error bars]{andrews2013} from 
the Sloan Digital Sky Survey Data Release 7 \citep[SDSS DR7]{abazajian2009}; the offset between model and data 
might be due to \textit{(i)} the assumed IMF when fitting the observed galaxy spectral energy distribution (SED), which mostly affects the 
galaxy stellar mass estimates, and/or \textit{(ii)} the assumed calibration in the chemical abundance measurement, 
which can strongly affect the O/H abundances 
(see, for example, \citealt{kewley2008,belfiore2017}).   
Our redshift evolution of the stellar mass--O/H relation is slightly weaker than in \citet{taylor2016}, which is due to the failed SN scenario assumed in this work. 
We note that the feedback from active galactic nuclei (AGNs) is not included in our simulation, however
the effect of AGN feedback has been shown to be negligible for this relation \citep{taylor2015a,taylor2015b}. 

\begin{figure}
\centering
\includegraphics[width=8.0cm]{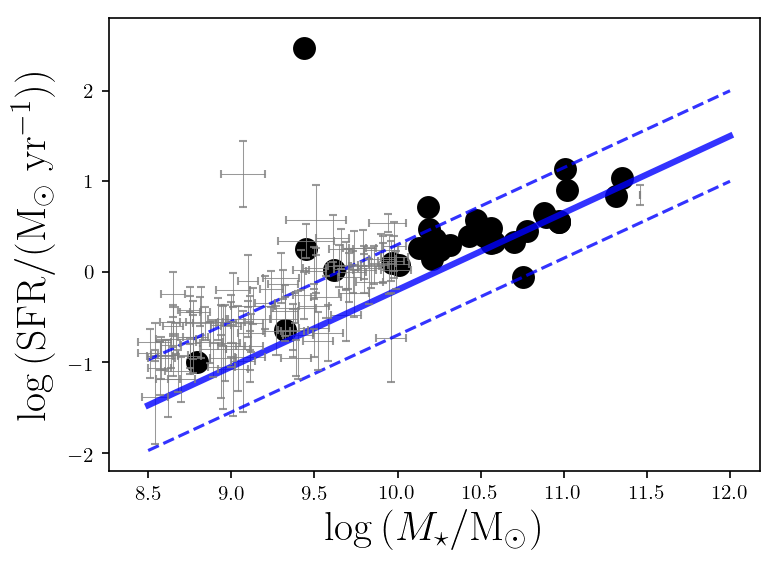} 
\caption{ {The $\log\Big(M_{\star}/\text{M}_{\sun}\Big)$--$\log\Big(\text{SFR}/(\text{M}_{\sun}\,\text{yr}^{-1})\Big)$ relation as predicted 
at redshift $z=0$ for our galaxy catalogue (black points) and compared with the observed data of \citet[grey points with error bars]{cicone2017} 
and the observed average relation of \citet[blue solid line]{belfiore2017b} with the corresponding $1\sigma$ scatter (blue dashed lines). } 
}
\label{fig:mass_sfr}
\end{figure}

Since the galaxy stellar mass strongly correlates with the average galaxy ISM metallicity, 
there is also a correlation between stellar mass and the average N/O ratio in the ISM, which is 
mostly due to the secondary behaviour 
of the N stellar yields from massive stars and AGB stars. In this way, we can explain the observed stellar mass--N/O relation of \citet{andrews2013} in the redshift range 
$0.027 < z < 0.25$, although 
there is an offset between model and data.

Finally, because we weight the global O and N abundances with the SFR, we show that the SFRs of our simulated galaxies 
are in good agreement with observations. In Figure \ref{fig:mass_sfr}, the predicted global SFR--$M_{\star}$ relation 
at redshift $z=0$ in our galaxy catalogue (black points) 
is compared with the observed data from \citet[grey points with error bars]{cicone2017} and the observed average relation from \citet{belfiore2017b}, which 
is consistent with the \citet{renzini2015} relation. Although 
the scatter in the predicted average SFRs is large, our simulation qualitatively agrees with observations. 
Finally, we note that the red sequence cannot be produced without the feedback from AGNs (\citealt{taylor2015a,taylor2015b}). 

\subsection{N/O evolution in a cosmological context} \label{sec:cosmological_context}

Since the global N/O--O/H relation in Figure \ref{fig:no_oh_all} is primarily driven by the mass-metallicity of galaxies, one may expect some environmental dependence.
Although the environmental dependence of mass-metallicity relation is
seen in some observational data \citep{ellison2009}, it is not so clear in other works \citep[e.g.,][]{kacprzak2015,pilyugin2017}. 

\begin{figure}
\centering
\includegraphics[width=8.0cm]{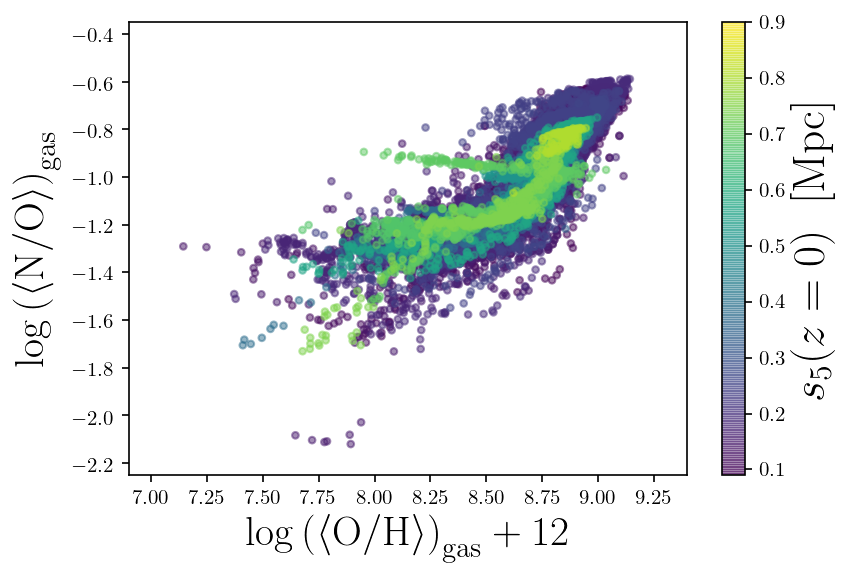} 
\caption{Same as Figure \ref{fig:no_oh_all} but coloured with the 5th nearest distance $s_{5}$ at $z=0$ to show the enviromental dependence. 
Higher values of $s_{5}$ indicate less dense environments. }
\label{fig:s5}
\end{figure}

In Figure \ref{fig:s5}, we show the effect of the environment on the evolution of galaxies in the N/O--O/H diagram.
As the indicator, we use the distance to the fifth nearest halo identified in our cosmological simulation at $z=0$, which represents the large-scale structures of galaxies very well (see Figure 5 of \citealt{taylor2017}).
High values of $s_{5}$ for a given DM halo (which can 
typically be as high as $\approx 0.9\,\text{Mpc}$) indicate relatively low densities of galaxies in the local environment. 
The various points in Figure \ref{fig:s5} represent the redshift evolutionary tracks 
in the N/O--O/H diagram as followed by all the galaxies in our catalogue 
with the colour coding indicating the $s_{5}$ index. 
Note that the galaxies are over-plotted in the order of $s_{5}$. 

By visually comparing Figures \ref{fig:no_oh_all} and
  \ref{fig:s5}, it can be seen that
the environmental dependence is much weaker than the redshift evolution; all galaxies follow the same N/O--O/H relation, indicating that the galaxy chemical evolution is self-regulated even with different SFHs.
There may be a small environmental dependence seen, where the
galaxies in the densest regions (with the lowest $s_{5}$ values) can reach 
higher average N/O ratios by the present time  at any fixed O/H abundance. 
We also find that these galaxies in the densest regions tend to show a larger scatter of evolutionary tracks in the N/O--O/H diagram.
This may indicate that not only the star formation efficiency (i.e., O/H) but also chemical enrichment timescale (i.e., N/O) may be different depending on the environment.
We should note, however, that this needs to be studied more with a large volume of simulations. 

\begin{figure}
\centering
\includegraphics[width=8.0cm]{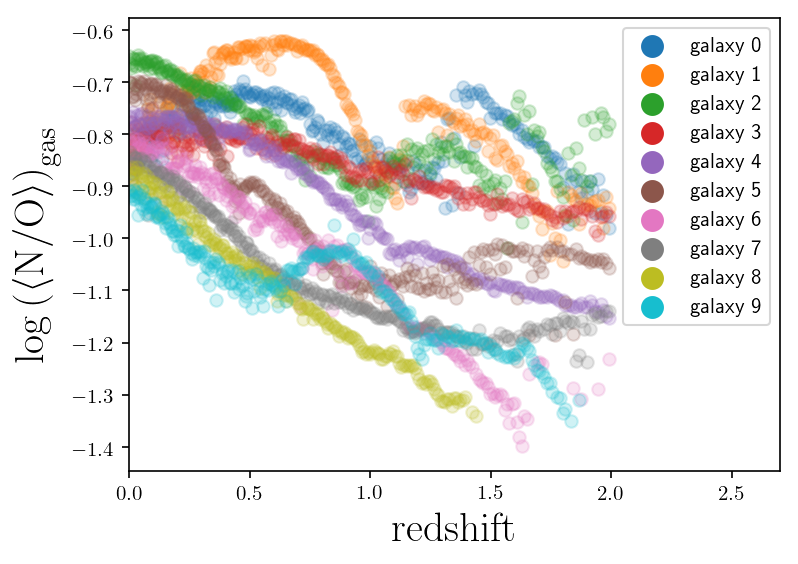} 
\caption{ The predicted redshift evolution of the average SFR-weighted N/O ratios within our ten reference galaxies in Figs. \ref{fig:gal_im_star} and \ref{fig:age_distribution}.  }
\label{fig:no_redshift}
\end{figure}

In contrast, there is a variation in the N/O evolution depending on the galaxy SFH. 
In Figure \ref{fig:no_redshift}, we show our predictions for the redshift evolution of the average SFR-weighted 
{$\log(\text{O/H})+12$ and} $\log(\text{N/O})$ ratios in the gas-phase of our ten reference galaxies (in Figs. \ref{fig:gal_im_star} and \ref{fig:age_distribution}), at redshifts $z\le2$ 
and for total stellar masses $M_{\star}\ge 10^{8}\,\text{M}_{\sun}$. 
First of all, we predict the average N/O ratios to increase, on average, in the 
galaxy ISM by the present; secondly, galaxies with a relatively smooth SFH, like galaxies $3$-$7$, and $8$, 
exhibit also a smooth increasing trend of the average gas-phase N/O ratio. 
Sudden bumps in the galaxy stellar mass growth history significantly affect the slope of the 
predicted N/O evolution, causing similar bumps in the predicted N/O evolution.

\begin{figure}
\centering
\includegraphics[width=8.0cm]{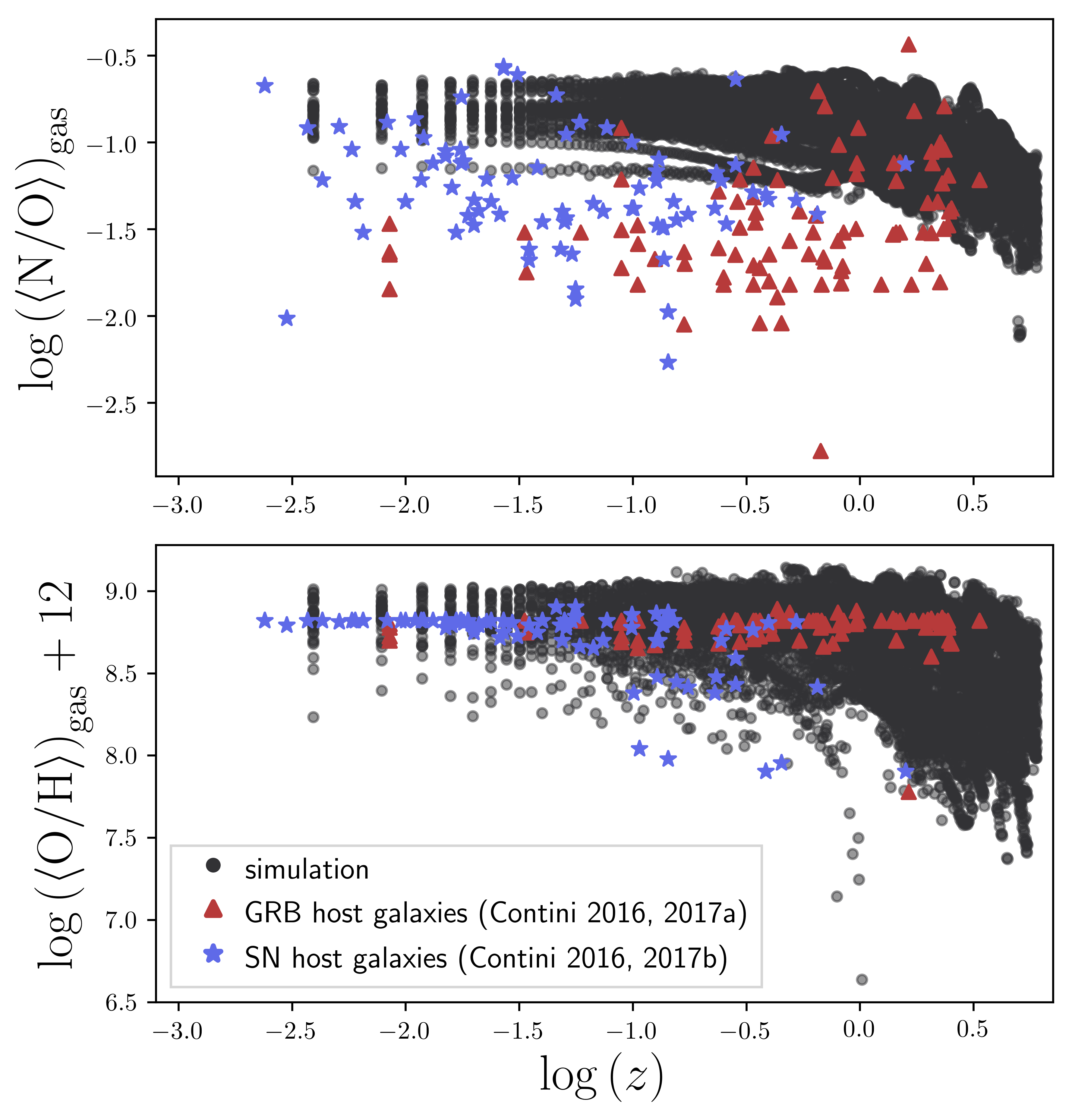} 
\caption{The predicted redshift evolution of N/O and O/H of all the galaxies in our catalogue (black circles in the top and bottom 
panels, respectively), comparing with 
available observations for SN host (blue stars) and GRB host (red triangles) galaxies (see the main text for more details).
}
\label{fig:contini}
\end{figure}

There are few observational studies in the literature which systematically 
attempted to measure N/O and O/H in galaxies at high redshifts; they mostly focused on AGN, gamma ray burst (GRB) or SN host galaxies, by 
making use of a detailed numerical modelling of the galaxy spectral energy distribution. Examples of such systematic studies are 
the series of works by \citet[but see also the previous works of the same author]{contini2015,contini2016,contini2017a,contini2017b,contini2018}, which adopted the 
\textsc{SUMA} numerical code\footnote{\url{http://wise-obs.tau.ac.il/~marcel/suma/index.htm}}, taking into account 
the combined effect of photoionisation and shocks \citep{contini1983,contini1986}. 
In Figure \ref{fig:contini}, we compare the predictions of our simulation for the redshift evolution of 
$\log(\text{O/H})+12$ and $\log(\text{N/O})$ (top and bottom panels, respectively) with the 
measurements in GRB and SN host galaxies (red triangles and blue stars, respectively). 
Our simulation tends to have lower O/H and thus higher N/O than in observations especially at high redshifts.
This is rather odd as it is the opposite from what we show in Figure \ref{fig:mass_met}.
This observational dataset may not be straight-forwardly comparable to our simulation; some spectra in the catalogue of \citet{contini2016,contini2017a,contini2017b} were taken in the very early phases after the SN explosion, 
before the SN ejecta disperse into the ambient ISM; this may eventually contaminate the abundance analysis, by leading to higher measured O/H abundances and hence lower N/O ratios. 
A similar tendency was reported for DLA systems, where GRB-DLA show higher metallicity than QSO-DLA at high redshifts \citep{cucchiara2015}. 
In future, comparisons with unbiased large samples of galaxies will
provide a more definitive test of our model predictions. 

\section{Conclusions} \label{sec:conclusions}

In this work, we have demonstrated that our model is capable of reproducing the observed 
increasing trend of N/O vs. O/H at high metallicity in the nearby star forming galaxies, by introducing failed SNe in our 
cosmological chemodynamical simulation. 

We have constructed a sample of $33$ star forming disc 
galaxies at redshift $z=0$, embedded within DM halos with virial mass in the range $10^{11} \le M_{\text{DM}} \le 10^{13}\,\text{M}_{\sun}$. 
We have analysed the detailed chemical evolution of the N and O abundances within ten reference galaxies of our catalogue,  
characterised by well distinct SFHs (Figure \ref{fig:age_distribution}). We have also 
shown how all $33$ galaxies 
in our catalogue evolve in the N/O--O/H, stellar mass--O/H and stellar mass--N/O diagrams, when considering SFR-weighted average abundances in 
the whole galaxy ISM. Our main conclusions can be summarised as follows.

\begin{enumerate}

\item The observed increasing trend of N/O at high O/H in the individual ISM regions of 
spatially-resolved star forming disc galaxies (we have referred to it as \textit{local} N/O--O/H relation in this work) 
can be explained as the consequence of metallicity 
gradients which have settled in the galaxy ISM, where the innermost regions possess both the highest O/H and the highest N/O ratios. 

\item The \textit{global} N/O--O/H relation when dealing with average abundances from the whole galaxy ISM is 
the consequence of an underlying mass-metallicity relation 
that galaxies obey as they evolve across the cosmic epochs. 
In this case, the predicted N/O--O/H relation is an average evolutionary trend which is followed by the chemical evolution tracks of all galaxies 
at almost any redshift. 

\item We do not find a strong environmental dependence but find that galaxies follow the same \textit{global} N/O--O/H relation independent of the environment ($s_5$).
However, galaxies in the densest environments at $z=0$ show a larger scatter along the relation, and thus can have higher N/O ratios at high O/H, than the galaxies in the least dense environments.

\item For both local and global relations, the increasing trend of N/O as a function of O/H 
is mainly due to the fact that N is mainly produced as secondary element 
at the expense of the C and O nuclei already present in the 
stars at their birth; the higher the initial stellar O/H abundance, the larger is the amount of 
synthesised N produced by stars. 

\item The average N/O ratios increase more rapidly in galaxies having SFHs concentrated at earlier cosmic epochs. Smooth stellar mass growth with time 
gives rise to smooth monotonic evolution of the average N/O ratios with redshift. Conversely, sudden bumps in the stellar mass growth history 
may also give rise to similar bumps in the $z$--N/O evolutionary tracks. Therefore, the redshift evolution of N/O in galaxies could be used to contrain the SFH of disc galaxies. 

\item We predict that the O/H and N/O gradients in the ISM of galaxies flatten -- on average -- as functions of time, in agreement with previous studies on 
the metallicity gradient evolution in disc galaxies (e.g. \citealt{kobayashi2011a,pilkington2012,gibson2013}); 
contextually, we predict also an inside-out growth of the galaxy as a function of time. 
In the very early ``protogalactic'' evolutionary stages, we predict highly scattered and overall flat abundance gradients; then, as the first series of 
stellar populations form, steep gas-phase abundance gradients soon
develop and then gradually flatten by the present time.

\end{enumerate}



\section*{Acknowledgments}
We thank an anonymous referee for his/her constructive comments and suggestions on the cosmological context. 
FV acknowledges funding from the UK 
Science and Technology Facility Council (STFC) through grant ST/M000958/1. 
This research has made use of the DiRAC high-performance computing (HPC) facility in the UK at Durham, 
supported by STFC and BIS; we also have made use of the University of Hertfordshire's HPC facility for the simulation analysis. 
CK acknowledges PRACE for awarding her access to resource ARCHER based in the UK at Edinburgh. 
Finally, we thank Volker Springel for providing the code 
\textsc{Gadget-3}, Francesco Belfiore for providing the observed data set and for many stimulating discussions, and Roberto Maiolino and 
Philip Taylor for fruitful discussions.

\end{document}